\documentstyle[12pt,cite,epsfig]{article}
\newcommand{\sect}[1]{\setcounter{equation}{0}\section{#1}}
\newcommand{\subsect}[1]{\subsection{#1}}

\evensidemargin=\oddsidemargin
\newcommand{\be}{\begin{equation}}
\newcommand{\ee}{\end{equation}}
\newcommand{\bea}{\begin{eqnarray}}
\newcommand{\eea}{\end{eqnarray}}
\newcommand{\p}{\partial}
\newcommand{\hi}{\hat{\i}}
\newcommand{\hj}{\hat{\j}}

\begin{document}
\renewcommand{\thefootnote}{\fnsymbol{footnote}}
\begin{titlepage}
\begin{flushright}
ROM2F/2004/15\\
IP/BBSR/2004-12\\
IMSc/2004/05/23\\
hep-th/0405124\\
\end{flushright}
\vskip .5in
\begin{center}
{\large\bf Brane Solutions with/without Rotation in PP-wave Spacetime}
\vskip .5in
{\bf Rashmi R. Nayak}$^a$\footnote{e-mail: {\tt rashmi@iopb.res.in}},
{\bf \,\,\,Kamal L. Panigrahi}$^b$\footnote{e-mail: {\tt
    Kamal.Panigrahi@roma2.infn.it}, INFN fellow},
{\bf \,and \,\,Sanjay Siwach}$^c$\footnote{e-mail: {\tt
    sanjay@imsc.res.in}}\\
\vskip .2in
{}$^a${\it Institute of Physics, Bhubaneswar 751 005, INDIA}\\
\vskip .2in
{}$^b${\it Dipartimento di Fisica, Universita' di Roma ``Tor Vergata"\\
INFN, Sezione di Roma ``Tor Vergata", Via della Ricerca Scientifica 1\\ 
00133 Roma, ITALY}
\vskip .2in 
{}$^c${\it Institute of Mathematical Sciences, CIT Campus, Taramani,\\
 Chennai, 600 113, INDIA}
\vspace{.3in}
\begin{abstract}
\vskip .2in
\noindent
We present two classes of brane solutions in pp-wave spacetime.
The first class of branes with a rotation parameter are constructed in an
exact string background with NS-NS and R-R flux. 
The spacetime supersymmetry is analyzed by solving 
the standard Killing spinor equations and is shown to preserve the 
same amount of supersymmetry as the case without the rotation. 
This class of branes do not admit regular horizon.
The second class of brane solutions
are constructed by applying a null Melvin twist to the
brane solutions of flat spacetime supergravity. 
These solutions admit regular horizon. We also comment on some thermodynamic 
properties of this class of solutions.  
 
\end{abstract}
\end{center}
\vfill

\end{titlepage}
\setcounter{footnote}{0}
\renewcommand{\thefootnote}{\arabic{footnote}}

\sect{Introduction}

Study of string theory in plane wave (or pp-wave) background has
drawn lots
of attention in the last couple of years, in search of establishing
AdS/CFT like dualities. These backgrounds can be seen as  
a small deformation of ten dimensional Minkowski spacetime\cite{BFHP}. 
Plane wave (or pp-wave) spacetime qualifies, the most, for analyzing 
certain issues of quantum gravity and give a consistent background 
for studying string theory in light-cone gauge. These backgrounds are obtained
by applying Penrose-Guven limit on $AdS_p \times S^q$ type of geometries
and also from the near horizon geometries of various supergravity 
solutions in diverse dimensions. Of particular interest, is the maximally
supersymmetric pp-wave background which is obtained from the near horizon
geometry of coincident D3-branes in ten-dimensional spacetime 
in the Penrose limit. String theory in this background is exactly solvable 
in light-cone gauge and is shown to be dual to  
${\mathcal N} = 4$ super Yang-Mills theory in large R charge sector\cite{BMN}.
The PP-wave/CFT dualities have been 
analyzed (See\cite{sadri} for the updated references on this subject), 
and some speculations have been made regarding the `holography' 
in plane wave backgrounds. Whereas the above issues
are slightly more clear in backgrounds with NS-NS 3-form flux
\cite{KP,mathur,strom,GN,KD,BDKZ}
(e.g. the Nappi-Witten backgrounds and $AdS_3\times S^3$ spacetime),
they are not very profound in the case of maximally supersymmetric plane wave
background with R-R flux. 

Plane wave spacetime 
in the presence of non-constant flux is also an interesting background
to study string theory, as it provides examples of integrable models on
the worldsheet\cite{MM,RT1}. These backgrounds can also be interpreted as 
the deformation of the flat spacetime and are supported by null matter fields. 
The corresponding worldsheet theory 
is described by the nonlinear sigma model\cite{MM,RT1,Kim,Bonelli}. 
and represent the nontrivial
examples of interacting theories in light-cone gauge. The pp-wave backgrounds
with non-constant 3-form NS-NS ($H_3$) and R-R ($F_3$) flux 
do not admit, in contrast
to their 5-form R-R flux ($F_5$) counterpart, the linearly realized 
`supernumerary' killing spinors. Moreover these theories are closely 
related to the closed strings in a constant 
magnetic field, in the presence of antisymmetric tensor fields and 
a non-trivial metric\cite{RT94}, which also provides an example 
of $\alpha'$ exact string theory background. These backgrounds are known to be 
homogeneous plane wave backgrounds and string theory in these type
of spacetime has been analyzed in great detail \cite{Shahin,blau}.  

In recent years, several important aspects of 
string dualities have been revealed and D-branes have played important role
in these developments. While D-branes can be treated as the black branes
in supergravity theories, the effective field theory on the brane are
of super Yang Mills type theories. So the study of D-branes in various 
non trivial and non generic backgrounds with/without flux give 
ideas about the structures of string theory as well as the related 
gauge theories. D-branes in various curved background have
generated renewed interest in the context of plane wave background for
various reasons.
First, these nonperturbative objects are easily tractable in 
pp-wave background and second, the supergravity solutions
can be constructed with not much efforts as compared to its AdS counterpart. 
Various D-brane supergravity solutions
in maximally and less than maximally supersymmetric plane wave 
backgrounds have been analyzed in the past couple of 
years
\cite{kumar1,sken,bain,kumar2,kamal1,rashmi1,alday,sk,OPS,Brecher1,NP,HNP}. 
In this context various attempts have also been made 
in finding out black hole/brane solutions with regular horizon in
plane wave space time. The analysis of \cite{LPV,OPS} shows
strong evidence in favour of the non-existence of horizon in the spacetime
with covariantly constant and null Killing vectors, proposed in
\cite{rangamani1}. Moreover, the brane solutions seem to be singular.
In \cite{aki}, however, a consistent
method for obtaining a black string solution with regular horizon has 
been discussed, which relies on a solution generating technique\cite{alish} 
known as null Melvin twist (NMT). This particular mechanism, 
transforms a flat spacetime to a plane wave spacetime. So the natural 
guess would be to start with a black brane
solution in flat spacetime and apply NMT to obtain a solution in plane wave 
spacetime which preserves horizon. As proved in \cite{aki}, 
this particular transformation
indeed preserve the horizon and that the area of the horizon remains the same 
even after the NMT. 

Motivated by the recent interest in finding out exact string
backgrounds and the classical solutions of branes and their
bound states, in various nontrivial backgrounds with/without flux,
in this paper, we present some Dp and Dp-Dp' branes
in homogeneous  pp-wave background with non-constant flux
and with a rotation parameter. This class of solutions are
seen not to admit a regular horizon. We also examine another class 
of D-branes which are obtained by applying a solution generating technique 
known as null Melvin twist.
These class of solutions do keep the asymptotic of the spacetime
as that of the  plane waves but don't give rise to the 
null matter content of the theory. The rest of the paper, is 
organized as follows. In section-2, after a small digression
for the discussion of the homogeneous plane wave space time,
we present the classical solutions of some Dp as well as Dp-Dp'
branes in this background 
with an explicit inclusion of the 
rotation parameter. We keep the fluxes completely general
and show that they solve supergravity field equations.
In section-3, we analyze the supersymmetry 
of the background and the branes in this background by solving the
Killing spinor equations. The solutions of the Killing spinor
equations are shown to constrain the structure of the 3-form
fluxes. We also make some remarks on the properties of horizon
of these brane solutions.
In section-4, we present classical solutions of black branes
in asymptotically plane wave spacetime by applying the null Melvin
twist on the non-extremal brane solutions of flat spacetime supergravity.
We also compute the horizon area and temperature of these black branes.    
In section-5, we conclude with some remarks and discussions.
  
\sect{Branes with rotation in  pp-wave background}
\subsect{The Background}
As a warm up exercise, below, we recapitulate few basic facts
about the homogeneous plane wave background discussed in
\cite{blau}, which will be helpful in fixing the notations etc. 
The most general null Brinkmann metric in d-dimensions with flat 
transverse space is given by:
\bea
ds^2 = 2du dv + {\mathcal H}(u, x) du^2 + 2 A_i (u, x)dx^i du + dx^i dx_i
\eea
Exact string backgrounds, with this metric, can be constructed by
switching on the appropriate field strengths and the dilaton:
\bea
{\mathcal B}^{NS}_{i u} = B_i (u, x),\>\>\>\>\>\> \phi = \phi (u).  
\eea  
For the above ansatz, the one-loop conformal invariance, or in other words,
the type II supergravity field equations give the following conditions:
\bea
&&-{1\over 2} \Box {\mathcal H} + \partial_u \partial^i A_i
+ {1\over 4} {\mathcal F}_{ij}{\mathcal F}^{ij} -
{1\over 4}{H}_{ij}{H}^{ij} + 2 \partial^2_u \phi = 0, \cr
& \cr
&& \partial^i {\mathcal F}_{ij} = 0, \>\>\> \partial^i {H}_{ij} = 0,
\label{exact}
\eea
where ${\mathcal F}_{ij} = \partial_i A_j - \partial_j A_i$ and 
${H}_{ij} = \partial_i B_j - \partial_j B_i$. In principle, 
the general solutions to these equations do not define an exact 
background. Some of the special cases where it does, has been 
discussed in \cite{blau}.

In the present paper, we shall be interested in the supergravity background 
with the metric, NS-NS 2-form (${\mathcal B}$) and R-R two form 
(${\mathcal B'}$) (and a constant dilaton): 
\bea
ds^2 &=& 2 du dv + {\mathcal{H}}(x_i) du^2 + 2 J A_i (x_i)dudx^i 
+ \sum_{i=1}^4{(dx^i)}^2 + \sum_{a=5}^8{(dx^a)}^2, \cr
& \cr
{\mathcal B} &=& b_i (x_i) du\wedge dx^i,\>\>\>H_3 = H_{i j}(x_i) \>\> 
du\wedge dx^i\wedge dx^j, \cr
& \cr
{\mathcal B'} &=& JA_i (x_i) du\wedge dx^i, \>\>\> 
F_3 = J {\mathcal F}_{ij}(x_i) \> du\wedge dx^i \wedge dx^j, 
\label{bkgrd}
\eea
where $H_3$ and $F_3$ are the field strengths associated with the
${\mathcal B}$ and ${\mathcal B'}$ respectively: 
${\mathcal F}_{ij} = \partial_i A_j - 
\partial_j A_i$ and $H_{i j} = \partial_i b_j - \partial_j b_i$.
$J$ is an arbitrary constant parameter. Note that we are considering the 
case in which the above metric functions are independent of the light-cone 
time $u$. We keep the most general form of the NS-NS and R-R field strengths,
but the restrictions on them would be imposed by the requirement of
supersymmetry, as we will see in the subsequent analysis.
Few remarks regarding the structure of $A(x^i)$
are in order now. If we restrict:
$JA_i dx^i du =  J \epsilon_{ij} x^i dx^j du$, where 
$\epsilon_{12} = \epsilon_{34} = 1$, then this can be interpreted
as the rotation in the $x^i$ space of solutions, parametrized by $J$
\footnote{Similar analysis have also been performed in the context
of closed strings in the presence of magnetic field, and  
proved to be conformally invariant background\cite{RT94}}.
That would further restrict the background fields turned on, which 
in turn play an important role in the analysis of 
Killing spinor equations. We will come back to this issue later on.      

To be a consistent solution of supergravity, the above ansatz should 
be supplimented by the constraints:
\bea
&&\Box^{(i)}{\mathcal H}(x_i) = - {(\partial_{i} b_{j})}^2 , 
\>\>\ \partial^k {\mathcal F}_{ki} = 0,\>\>\ \partial^k H_{k i} = 0.
\label{cond}
\eea 

We have
checked that the solution (\ref{bkgrd}) supplimented by the conditions
(\ref{cond}) satisfy all the type IIB field equations and the Bianchi
identities. We shall be interested
in this background for the subsequent analysis of the present paper. 
As can be seen from the metric (\ref{bkgrd}) that switching off the 
gauge field $(A)$ we get back to the pp-wave metric of \cite{RT1} 
with non-constant NS flux. 

\subsect{Supergravity solutions}

In this section, we present classical solutions of branes in the 
above background supported by NS-NS ($H_3$) and R-R ($F_3$)
flux in the
transverse direction of the branes. We start by writing down the
supergravity solution of $N$ D-strings lying on top of each other 
in this background. The metric, the dilaton, and the field strengths 
of such a configuration is given by:
\bea
ds^2 &=& f^{-{1\over 2}}_1\Big(2 du dv + {\mathcal{H}}(x_i) du^2 
+ 2 J A_i(x_i) dudx^i\Big) + f^{1\over 2}_1 \Big(\sum_{i=1}^4{(dx^i)}^2 \cr
& \cr
&+& \sum_{a=5}^8{(dx^a)}^2\Big), \cr
& \cr
H_3 &=& H_{ij} (x_i)\>\> du\wedge dx^i\wedge dx^j, \>\>
e^{2\phi} = f_1, \cr
& \cr
{\mathcal B'} &=& (f^{-1}_1 -1)~du\wedge dv 
+ {J\over f_1} A_i (x_i)~du\wedge dx^i,
\label{D-string}
\eea
where $f_1 = 1 + {{Q_1}\over{r^6}}$ is the harmonic function in the transverse
space of the D-string and ${\mathcal B'}$ is the Ramond-Ramond potential. 
We have
checked that the above solution solves type IIB field equations provided
the constraints (\ref{cond}) are also imposed. Similarly, a 
D5-brane solution is given by:
\bea
ds^2 &=& f^{-{1\over 2}}_5\Big(2 du dv + {\mathcal{H}}(x_i) du^2 
+ 2 J A_i(x_i) dudx^i + \sum_{a=5}^8{(dx^a)}^2 \Big) \cr
& \cr
&+& f^{1\over 2}_5 \sum_{i=1}^4{(dx^i)}^2, \cr
& \cr
H_3 &=& H_{ij} (x_i)\>\> du\wedge dx^i\wedge dx^j, \>\>
e^{2\phi} = f^{-1}_5, \cr
& \cr
F_3 &=& J {\mathcal F}_{i j} (x_i)~du\wedge dx^i \wedge dx^j, \>\>\> 
F_{ijk} = \epsilon^l_{ijk}\partial_l f_5,
\label{D5-brane}
\eea
where $f_5 = 1 + {{Q_5}\over{r^2}}$ is the harmonic function in the 
transverse 4-space. $F_3$
and $F_{ijk}$ are the Ramond-Ramond field strengths. We have once again
checked that the above ansatz solves type IIB field equations provided
the constraints (\ref{cond}) are also imposed. 
The above solutions reduce to those presented in \cite{sk} for $J = 0$. 
Other Dp-brane (for
$p\ge 2$)solutions can also be constructed first by smearing 
$a = x^5,..x^8$ directions and then by applying T-dualities along those. 

Now we present the classical solution of D1-D5 system, as an example of
Dp-D(p+4) brane bound state in the background (\ref{bkgrd}). The metric,
dilaton, NS-NS  and R-R fields of such a configurations are 
given by:
\bea
ds^2 &=& (f_1 f_5)^{-{1\over 2}}\Big(2 du dv + {\mathcal{H}}(x_i) du^2 
+ 2 J A_i(x_i) dudx^i \Big)+ {({f_1\over f_5})}^{1\over 2}
\sum_{a=5}^8{(dx^a)}^2 \cr
& \cr
&+& (f_1 f_5)^{1\over 2}\sum_{i=1}^4{(dx^i)}^2, \cr
& \cr
H_3 &=& H_{ij} (x_i)\>\> du\wedge dx^i\wedge dx^j, \>\>
e^{2\phi} = {f_1 \over f_5}, \cr
& \cr
{\mathcal B'} &=& (f^{-1}_1 -1)~du\wedge dv + {J\over f_1} A_i (x_i)~du
\wedge dx^i, \cr
& \cr 
F_{ijk} &=& \epsilon^l_{ijk}\partial_l f_5,
\label{d1-d5-brane}
\eea
where $f_{1,5} = 1 + {Q_{1,5} \over r^2}$ are the harmonic functions
of D1 and D5 brane in the common transverse 4-space. We have once again 
checked that the above ansatz solves type IIB field equations provided the
identities (\ref{cond}) are also imposed. This solution reduces to the
D1-D5 brane solution of \cite{sk} for J = 0. We would like to point out 
that a similar solution has already been presented in \cite{Brecher1}.
However the choice of the background flux turned on to compensate the 
effect of ${\mathcal H}$ is different here. 
The corresponding differential equations for ${\mathcal H}$  
is given by (\ref{cond}) and is independent of the parameter $J$ for 
all the solutions. The explicit solution can be read off from the reference 
\cite{OPS} for the branes presented here.
     
\sect{Supersymmetry}

The supersymmetry variation of dilatino and 
gravitini fields of type IIB supergravity in ten dimensions, 
in string frame, is given by \cite{schwarz,fawad}:
\begin{eqnarray}
\delta \lambda_{\pm} &=& {1\over2}(\Gamma^{\mu}\partial_{\mu}\phi \mp
{1\over 12} \Gamma^{\mu \nu \rho}H_{\mu \nu \rho})\epsilon_{\pm} + {1\over
  2}e^{\phi}(\pm \Gamma^{M}F^{(1)}_{M} + {1\over 12} \Gamma^{\mu \nu
  \rho}F^{(3)}_{\mu \nu \rho})\epsilon_{\mp},
\label{dilatino}
\end{eqnarray}
\bea
\delta {\Psi^{\pm}_{\mu}} &=& \Big[\partial_{\mu} + {1\over 4}(w_{\mu
  \hat a \hat b} \mp {1\over 2} H_{\mu \hat{a}
  \hat{b}})\Gamma^{\hat{a}\hat{b}}\Big]\epsilon_{\pm} \cr
& \cr
&+& {1\over 8}e^{\phi}\Big[\mp \Gamma^{\lambda}F^{(1)}_{\lambda} - {1\over 3!}
\Gamma^{\lambda \nu \rho}F^{(3)}_{\lambda \nu \rho} \mp {1\over 2.5!}
\Gamma^{\lambda \nu \rho \alpha \beta}F^{(5)}_{\lambda \nu \rho \alpha
  \beta}\Big]\Gamma_{\mu}\epsilon_{\mp},
\label{gravitino}
\end{eqnarray}
where we have used $(\mu, \nu ,\rho)$ to describe the ten
dimensional space-time indices, and hats represent the corresponding
tangent space indices. 

\subsection{Background supersymmetry}
Before analyzing the supersymmetry of the
rotating Dp and Dp-Dp' brane solutions in homogeneous plane wave background,
let's first discuss the supersymmetry of the background itself. 
The dilatino (\ref{dilatino}) and gravitino (\ref{gravitino})
variations impose nontrivial conditions on the spinor $\epsilon_{\pm}$.
First the dilatino variation gives:
\bea
\mp \Gamma^{\hat u\hat{\i}\hat{\j}}~H_{\hat{\i}\hat{\j}}\epsilon_{\pm} 
+ J \Gamma^{\hat u\hat{\i}\hat{\j}}~{\mathcal F}_{\hat{\i}\hat{\j}}
\epsilon_{\mp} = 0.
\eea 
Similarly, from the gravitini variations, we get the following conditions
on the spinors to have nontrivial solutions:
\bea
\delta \psi^{\pm}_{u}&\equiv& \left(\p_u + {J\over {4}} 
{\mathcal F}_{{\hi}{\hj}}\Gamma^{{\hi}{\hj}}
\mp {1\over 4} H_{{\hi}{\hj}}\Gamma^{{\hi}{\hj}}\right)\epsilon_{\pm} 
- {J\over 8}{\mathcal F}_{{\hi}{\hj}}\Gamma^{\hat u{\hi}{\hj}}
\Gamma^{\hat v}\epsilon_{\mp} = 0, \cr
& \cr
\delta \psi^{\pm}_{v}&\equiv& \p_v \epsilon_{\pm} = 0, \>\>\> 
\delta \psi^{\pm}_{a}\equiv \p_a \epsilon_{\pm} = 0, \>\>\>
\delta \psi^{\pm}_{i} \equiv \p_i \epsilon_{\pm} = 0.
\eea
In writing down the above supersymmetry variations we have made use of
the standard supersymmetry condition\footnote{this condition does not 
depend on the details of the pp-wave background that we are considering}
\bea
\Gamma^{\hat u}\epsilon_{\pm} = 0.
\label{standard}
\eea 
After imposing this condition the dilatino variation, above, is 
 satisfied. One notices that for the variations of the remaining
terms in the gravitino variation $\delta \psi^{\pm}_{u}$, for a constant
spinor $\epsilon_0$, we need to restrict the structure of the background 
flux ${\mathcal F}_{i j}$ and also $H_{i j}$. One such possibility has been
discussed in\cite{sk}. 
For the case: $F^{(3)}_{ u 1 2} = F^{(3)}_{u 3 4}$
and $H^{(3)}_{ u 1 2} = H^{(3)}_{u 3 4}$, with all other components
of ${\mathcal F}_{ij}$ and $H_{ij}$ set to zero,
we have to impose the condition:
\bea
(1 - \Gamma^{\hat 1\hat 2\hat 3\hat 4})\epsilon_{\pm} = 0.
\label{pp-wave}
\eea
So the amount of supersymmetry preserved, after imposing the above two
conditions, (\ref{standard}) and (\ref{pp-wave}) is 1/4 of the original one. 
This fact has also been shown
in \cite{herdeiro}. We would like to point out that the addition of
the the rotation J, does not destroy more supersymmetry compared to 
the case without J. So the natural guess would be that the fate of the
remaining supersymmetry in the presence of D-branes will be the same as
the case without J, that has been explained in\cite{sk,NP}. We will 
examine this fact by giving examples of D-branes that we have considered 
in the previous section. 
 
\subsection{D-brane supersymmetry}

In this section, we analyze the supersymmetry conditions for the
D-string (\ref{D-string}) and the D1-D5 brane bound state (\ref{d1-d5-brane})
solutions presented in the previous section. First, the dilatino variation 
equation for the D-string solution presented in (\ref{D-string}) gives:
\bea
\left(\Gamma^{\hat{\alpha}}\epsilon_{\pm} - \Gamma^{\hat u\hat v\hat{\alpha}}
\epsilon_{\mp}
\right) \p_{\hat{\alpha}} f_1 \mp {{f^{5\over 4}_1}\over 4} 
\Gamma^{\hat u{\hi}{\hj}} 
~H_{{\hi}{\hj}} \epsilon_{\pm} + {J\over 4}f^{7\over 4}_1 
\Gamma^{\hat u{\hi}{\hj}}~{\mathcal F}_{{\hi}{\hj}}\epsilon_{\mp} = 0, 
\>\> \alpha=1,..,8.
\eea
On the other hand, the gravitini variations gives the following conditions
on the spinors to have nontrivial solutions:
\bea
\delta \psi^{\pm}_{u}&\equiv& \left(\p_u + {J\over {4}}f^{-{1\over 2}}_1 
{\mathcal F}_{{\hi}{\hj}}\Gamma^{{\hi}{\hj}}
\mp {1\over 4} H_{{\hi}{\hj}}~\Gamma^{{\hi}{\hj}}\right)\epsilon_{\pm} 
- {J\over 8}f^{1\over 2}_1{\mathcal F}_{{\hi}{\hj}}~
\Gamma^{\hat u{\hi}{\hj}}~\Gamma^{\hat v}\epsilon_{\mp} = 0, \cr
& \cr
\delta \psi^{\pm}_{v}&\equiv& \p_u \epsilon_{\pm} = 0, \cr
& \cr
\delta \psi^{\pm}_{i} &\equiv& \left(\p_i + {1\over 8}{{\p_i f_1}\over f_1}
\right)\epsilon_{\pm} = 0,\>\>\>
\delta \psi^{\pm}_{a} \equiv \left(\p_{a} 
+ {1\over 8}{{\p_{a} f_1}\over f_1} \right)\epsilon_{\pm} = 0. 
\eea
In writing down the above gravitini variations, we have made use of the
D-string supersymmetry condition:
\bea
\epsilon_{\pm} - \Gamma^{\hat u\hat v}\epsilon_{\mp} = 0, 
\label{string-susy}
\eea
in addition to the necessary condition (\ref{standard}). By imposing 
(\ref{standard}), the dilatino variation is satisfied. The gravitini
variations $\delta \psi^{\pm}_{v}$, $\delta \psi^{\pm}_{i}$ and 
$\delta \psi^{\pm}_{a}$,
solve for the spinor: \\
$\epsilon_{\pm} = \exp\left(-{1\over 8}\ln f_1\right)
\epsilon^0_{\pm}$, with $\epsilon^0_{\pm}$ being a spinor
which can depend on the coordinate $u$, leaving
the following equations to have a nontrivial solution:
\footnote{since $\epsilon^0_{\pm}$ is a function of $u$ only and both
${\mathcal F}_{ij}$ and $H_{ij}$ are functions of $x^i$ only}:
\bea
\left( {J\over {4}}f^{-{1\over 2}}_1 
{\mathcal F}_{{\hi}{\hj}}\Gamma^{{\hi}{\hj}}
\mp {1\over 4} H_{{\hi}{\hj}}~\Gamma^{{\hi}{\hj}}\right)\epsilon^0_{\pm} 
- {J\over 8}f^{1\over 2}_1{\mathcal F}_{{\hi}{\hj}}~
\Gamma^{\hat u{\hi}{\hj}}~\Gamma^{\hat v}\epsilon^0_{\mp} = 0, 
\eea
and
\bea
\p_u \epsilon^0_{\pm} = 0.
\eea
One can see that the existence of the solution to the above
equations can be obtained by restricting the form of the functions
${\mathcal F}_{ij}$ and $H_{ij}$. By making the choice: 
$F_{u12} = F_{u34}$ and $H_{u12} = H_{u34}$, we get an additional
condition on the spinors $\epsilon^0_{\pm}$:
\bea
(1-\Gamma^{\hat 1\hat 2\hat 3 \hat 4})\epsilon^0_{\pm} = 0.
\label{nes}
\eea    
Coming back to the counting of the surviving supersymmetry (which are
of `standard type' only), it is easy to see that the D-string solution
(\ref{D-string}), after imposing the conditions (\ref{string-susy}), 
(\ref{nes})
along with (\ref{standard}), preserves 1/8 of the original supersymmetry.
Therefore, even with the addition
of the rotation J, the D-string supersymmetry remains the same as 
in the non-rotating case, discussed in \cite{sk}. One can also 
show in a similar way that the D5-brane solution presented in (\ref{D5-brane})
preserves 1/8 of the supersymmetries.

Now let's analyze the supersymmetry of the D1-D5 brane bound state 
solution presented in (\ref{d1-d5-brane}). First the dilatino variation
gives:
\bea
\delta \lambda_{\pm} &\equiv& \left(\Gamma^{{\hi}} \epsilon_{\pm} 
- \Gamma^{\hat u\hat v{\hi}}
\epsilon_{\mp}\right) {{{f_{1,{\hi}}}} \over {f_1}} - 
\left(\Gamma^{{\hi}} \epsilon_{\pm} - {1\over 3!}
\epsilon^{\hi}_{{\hj}\hat k\hat l}\Gamma^{{\hj}\hat k\hat l}
\epsilon_{\mp}\right) {{{f_{5,{\hi}}}} \over {f_5}} \mp 
{(f_1 f_5)}^{1\over 4}\Gamma^{\hat u{\hi}{\hj}} H_{{\hi}{\hj}}
\epsilon_{\pm} \cr
& \cr
&+& J \left({f^3_1\over f_5}\right)^{1\over 4}\Gamma^{\hat u{\hi}{\hj}} 
{\mathcal F}_{{\hi}{\hj}}\epsilon_{\mp} = 0.  
\eea
The gravitino variations are very similar to those presented for the
D-string case. Therefore we skip the detailed expressions for those.
After imposing respectively the D-string and the D5-brane supersymmetry 
conditions: 
\bea
\Gamma^{{\hi}} \epsilon_{\pm} - \Gamma^{\hat u\hat v{\hi}} \epsilon_{\mp} = 0,
\label{D1-susy}
\eea
and 
\bea
\Gamma^{{\hi}} \epsilon_{\pm} - {1\over 3!}
\epsilon^{\hi}_{{\hj}\hat k\hat l}\Gamma^{{\hj}\hat k\hat l}
\epsilon_{\mp} = 0,
\label{D5-susy}
\eea 
along with the standard supersymmetry condition (\ref{standard}), 
the dilatino variation is
fully satisfied. The gravitini variation equations, however, would
require, an additional condition:
\bea
(1-\Gamma^{\hat 1\hat 2\hat 3\hat 4})\epsilon^0_{\pm} = 0,
\label{ness}
\eea
for the existence of a constant spinor ($\epsilon^0_{\pm}$) solution 
similar to that presented for the D-string case. Let's now count the
amount of supersymmetry preserved after imposing all these conditions.
First of all, $\Gamma^{\hat u} \epsilon_{\pm} = 0$ breaks half of
the supersymmetries. The fate of the remaining supersymmetries can be
found out by examining the conditions (\ref{D1-susy})-(\ref{ness}).
It is easy to see however that they indeed are only two independent
conditions on the spinor $\epsilon$. So the D1-D5 solution presented
in (\ref{d1-d5-brane}) preserves 1/8 of the supersymmetries.

Few remarks are in order now. As can be seen from the supersymmetry 
analysis of the D-branes in the pp-wave background, there is always
a 'decoupling' between the standard D-brane supersymmetry conditions
and the supersymmetry condition imposed by the pp-wave (which in turn
comes from the light-cone gauge fixing). 
We would like to note that
the branes that we are considering here are longitudinal branes\cite{bain}
(all the light-cone directions fall into the worldvolume directions
of the branes and other pp-wave directions are transverse to the
brane). 

\subsect{Horizon/No horizon}

The plane wave spacetime with covariantly constant and null Killing 
isometries, as such do not admit horizon \cite{rangamani1}.
Relaxing the covariant constancy condition raised some hope that there might 
exist horizon in spacetime admitting null killing isometry only (e.g. p-brane 
solutions in pp-wave spacetime)\cite{LPV}. 
However, the analysis of \cite{LPV,OPS} shows strong evidence in favour of 
the non-existence of  regular horizon for p-branes in pp-wave spacetime.

To examine the issue that the solutions presented in previous section admit 
regular horizon or not, one would like to see how the curvature tensors behave 
in the near horizon limit. Without addition of the rotation term, it has been 
noticed earlier that the potential divergent quantities are the components of 
the Riemann tensor \cite{brecher2,OPS}. An invariant measure of the divergence 
are the Riemann tensor components as measured in an orthonormal frame. 
A natural
choice for it is the parallel transported frame as emphasized in 
\cite{brecher2}. One can show that in the parallel transported frame some of 
the Riemann tensors diverge in the near horizon geometry, thereby showing the 
appearance of pp-curvature singularities\cite{brecher2,OPS}. Hence their 
doesn't exist the regular event horizon. Generalization to the non-extremal 
solutions along the lines of  \cite{LPV,OPS} doesn't improve the situation. 
It is not difficult to see that the addition of rotation term doesn't 
change the situation in both extremal and non-extremal cases.

However it has been argued that the pp-singularities close-off  of spacetime 
near the horizon\cite{brecher2}, thereby acting as the boundary of the spacetime and if one accepts this interpretation, the issue that the horizon is a 
singular surface becomes less important and the solutions can be thought of 
well behaved.

\sect{PP-wave branes from flat spacetime brane solutions}

We start by writing down the most general non-extremal   
p-branes in ten dimensions. The metric, dilaton and 
the field strengths of such non-extremal p-branes in ten dimensional spacetime 
is given by:
\bea
ds^2 &=& H^{-{1\over 2}}_p\Big(-f(r)dt^2 + \sum^p_{\alpha=1}{(dy^{\alpha})}^2
\Big) + H^{1\over 2}_p(f^{-1}(r)dr^2 + r^2 d\Omega^2_{\tilde d  + 1}),\cr
e^{\phi} &=& H^{{3-p}\over 4},\>\>\>   E^{(p+1)} = \tilde Q {f(r)\over H_p} 
dt\wedge dy^1\wedge dy^2 \wedge...\wedge dy^p,\cr
& \cr 
H_p &=& 1 + {Q_p\over r^{\tilde d}},\>\>\> f(r) = 1 - {\mu \over r^{\tilde d}}.
\label{nDp}
\eea
Where $H_p$ is the harmonic function for the $p$-branes which satisfies the
Green function equation in the transverse $\tilde d + 2 = 9-p $ space 
and $f(r)$ is the nonextremal parameter. 

Now we apply the solution generating technique, NMT as described 
in \cite{aki} on (\ref{nDp}) to generate black $p$-branes
in the plane wave spacetime. The first step involves a
boost along $y_1$ (which is one of the isometry directions along the brane). 
The resulting metric and field strength of the boosted $p$-brane becomes:
\bea
ds^2 &=& H^{-{1\over 2}}_p\Big(-\hat{K}^{-1}(r) f(r) d\tilde t^2
+ \hat{K}(r)[d\tilde y_1 + A(r)d\tilde t]^2 
+ \sum^{p}_2 {(dy^{\alpha})}^2\Big)\cr
& \cr
&+& H^{1\over 2}_p(f^{-1}(r)dr^2 + r^2 d\Omega^2_{\tilde d + 1}),\cr
& \cr
e^{\phi} &=& H^{{3-p}\over 4},\>\>\> E^{(p+1)} = \tilde Q \frac{f}{H_p}~
\gamma d\tilde t\wedge
d\tilde y^1\wedge dy^2 \wedge .....\wedge dy^p,
\label{nDpb}
\eea
where:
\bea
dt &=& \cosh \gamma d\tilde t - \sinh \gamma d\tilde y^1, \>\>\> 
dy^1 = - \sinh \gamma d\tilde t + \cosh \gamma d\tilde y^1,\cr
& \cr
\hat K(r) &=& 1 + \frac{\hat Q}{r^{\tilde d}},\>\>\>A(r) = -\frac{Q}
{r^{\tilde d}} {\hat K}^{-1},\>\>\>\hat Q = \mu \sinh^2 \gamma,\>\>\>
Q = \mu \sinh \gamma \cosh \gamma.
\eea

The second step involves a T-duality along $\tilde y^1$. The resulting solution 
is a boosted $(p-1)$-brane in ten dimensional spacetime with the following 
form of the metric, dilaton and the other fields:
\bea
ds^2 &=& H^{-{1\over 2}}_{p-1}\Big(-\hat{K}^{-1}(r) f(r) d\tilde t^2 
+ \sum^{p}_{\alpha =2}
{(dy^{\alpha})}^2\Big) + H^{1\over 2}_{p-1}\Big(\hat K^{-1}(r) 
{d\tilde y^1}^2\cr
& \cr
&+& f^{-1}(r) dr^2 + r^2 d\Omega^2_{\tilde d + 1}\Big),\cr
& \cr
e^{\phi} &=& {H^{{3-(p-1)}\over 4}_{p-1}\over \sqrt{{\hat K}(r)}},\>\>\>\> 
B = A(r)~dt\wedge d\tilde y^1, \cr
& \cr
E^{(p)} &=& \tilde Q \frac{f}{H_{p-1}}~d\tilde t\wedge dy^2 \wedge.....
\wedge dy^{(p-1)},
\eea

The next and the vital step of the NMT is to apply a twist 
$\sigma\rightarrow\sigma +2\alpha d\tilde y_1$, where the one 
form $\sigma$ is defined such that 
$d\Omega^2_{\tilde d+1} = \frac{1}{4}\sigma^2  + d\Sigma^2_{\tilde d}$. 
The spacetime, after this twist becomes:
\bea
ds^2 &=& H^{-{1\over 2}}_{p-1}\Big(-{\hat{K}^{-1}(r)}f(r) d\tilde t^2 
+ \sum^p_{\alpha = 2}   
{(dy^{\alpha})}^2\Big) + H^{1\over 2}_{p-1}\Big(\hat K^{-1}(r) 
{(d\tilde y^1)}^2\cr
& \cr
&+& f^{-1}(r)dr^2 
+ {1\over 4} r^2 (\tilde \sigma + 2\alpha d\tilde y^1)^2
+ r^2 d\Sigma^2_{\tilde d}\Big),\cr
& \cr
e^{\phi} &=& {H^{{3-(p-1)}\over 4}_{p-1}\over \sqrt{{\hat K}(r)}},
\>\>\>\> B = A(r)~dt\wedge d\tilde y^1, \cr
& \cr
E^{(p)} &=& \tilde Q \frac{f}{H_{p-1}}~d\tilde t\wedge dy^2 \wedge.....
\wedge dy^{(p-1)}.
\eea

Next one T-dualises back along $\tilde y^1$
to get back a $Dp$-brane solution and apply the inverse boost along 
$\tilde y_1$. The purpose of inverse boost is to cancel the boost charge 
as in step one. One gets the following configuration of metric and other fields:
\bea
ds^2 &=& H^{-{1\over 2}}_p\Big[ -\left({\hat{K}}^{-1}(r) f(r) \cosh^2 \gamma 
+ \frac{ (A(r)\cosh \gamma + \sinh \gamma)^2}{(\hat{K}^{-1}(r) 
+ r^2 \alpha^2)} \right) d t^2 \cr
& \cr
&+&2\Big(-\hat{K}^{-1}(r) f(r) \sinh \gamma \cosh \gamma +
{1\over {\hat{K}^{-1}(r) +  r^2 \alpha^2}}(A^2(r) \sinh \gamma \cosh \gamma\cr 
&\cr
&+& A(r)(\cosh^2 \gamma + \sinh^2 \gamma)
+ \sinh \gamma \cosh \gamma) \Big) d t d y_1 \cr
& \cr
&+& \left(-{\hat{K}(r)}^{-1} f(r) \sinh^2 \gamma + \frac{ (A\sinh \gamma 
b+ \cosh \gamma)^2}{(\hat{K}^{-1}(r) + r^2 \alpha^2)} \right) d{y_1}^2 
+ \sum^p_{\alpha=2} dy^2_{\alpha}\Big] \cr
& \cr
&+& H^{1\over 2}_p\Big[ {1\over f} dr^2 + {1\over 4}
{{ r^2 \hat{K}^{-1}(r)}\over{( \hat{K}^{-1}(r) + r^2 \alpha^2)}} 
\sigma^2 + r^2 d\Sigma^2_{\tilde d} \Big],\cr
& \cr
B &=&- {1\over 2}{{r^2 \alpha}\over {\hat{K}^{-1}(r) 
+ r^2 \alpha^2}}\Big[(\sinh\gamma + A(r) \cosh\gamma) dt\cr
& \cr
&+& (\cosh\gamma + A(r) \sinh\gamma)dy_1\Big]\wedge d\sigma \cr
& \cr
e^{\phi} &=& \frac{H_p^{\frac{(3-p)}{4}}}{{\sqrt{\hat{K}(r) 
\left(\hat{K}^{-1}(r) + r^2 \alpha^2\right)}}}, \>\>\>
E^{(p+1)} = \tilde Q \frac{f}{H_p} dt\wedge dy^1.....
\wedge dy^{p}, 
\eea

The final step is to take the limit $\alpha \rightarrow 0$ and 
$\gamma \rightarrow \infty$ while keeping $\beta= \frac{1}{2} \alpha e^\gamma$ 
fixed. The net effect of this step is to make the twist null hence the name 
null twist.

\bea
ds^2 &=& H^{-{1\over 2}}_p\Big[ -\frac{f(r)(1+\beta^2 r^2)}{k(r)} d t^2 
- \frac{2\beta^2 r^2 f(r)}{k(r)}d t d y^1 + \left(1-\frac{\beta^2 r^2}{k(r)}
\right){(d{y^1})}^2 \cr
& \cr
&+& \sum^p_{\alpha=2} dy^2_{\alpha}\Big]
+ H^{1\over 2}_p\Big[ {1\over f} dr^2 - {1\over 4}  
{{\beta^2 r^{4} (1-f(r))}\over{ k(r)}} \sigma^2 
+ r^2 d\Omega^2_{\tilde d +1} \Big],\cr
& \cr
e^{\phi} &=&\frac{H_p^{\frac{(3-p)}{4}}}{\sqrt{ k(r)}},\>\>\> ~~~~~
B =- {1\over 2}{{r^2 \beta}\over {k(r)}}\left( f(r) dt + dy^1 \right)\wedge 
\sigma,\cr
& \cr
E^{(p+1)} &=& \tilde Q \frac{f}{H_p} dt\wedge dy^1.....\wedge dy^{p},\>\>\>
~~~~~k(r) = 1 + {{\beta^2 \mu} \over r^{\tilde d -2}}.
\eea

The above solution reduces to the black p-brane in flat spacetime (\ref{nDp}) 
in $\beta \rightarrow 0$ limit and goes to the plane wave metric in 
ten-dimensions asymptotically. For $\beta = 0$, it is well known that the 
above metric admits a regular event horizon at 

\[
r_+ = \mu^{1/ \tilde d}.
\]

It was noticed by the authors of \cite{aki} in the context of black string 
solution in asymptotically plane wave spacetime that the solution obtained 
by NMT also admits regular horizon at $r_+$. It also remains true for the 
black brane solutions constructed above. Moreover the area of event horizon 
remains invariant under NMT i.e. 
it is independent of $\beta$. The area of event horizon 
(as measured in Einstein frame) is given by:
\bea
{\mathcal A} &=& \sqrt{\left(k(r)-\beta^2 r^2\right) H}~ r^{\tilde d + 1}
\Omega_{\tilde d + 1} \cr
&\cr
& = & \sqrt{ H(r_+)} \mu^{\frac{\tilde d + 1}{\tilde d}}.
\eea

One quarter of the area of event horizon measured in Plank units furnish 
the statistical entropy of the black branes according to the laws of 
black hole thermodynamics. Defining $ 1- \mu~ r^{-\tilde d} = \rho^2 $, 
and Euclidean time 
$ i\tau = t $, the metric can be put in the following form:
\bea
ds^2 = \frac{4 r_+^2}{{\tilde d}^2} H^{\tilde d /8}\left 
(d\rho^2 + \frac{{\tilde d}^2}{4 r_+^2} H^{-1}\rho^2 d\tau^2 + ...\right).
\eea

The temperature of the black brane solutions is given by the inverse 
periodicity of the Euclidean time:

\bea
T & = & \frac{\tilde d}{4\pi r_+} H^{-\frac{1}{2}}(r_+), 
\eea

which is also independent of $\beta$.

\sect{Discussion}

In this paper we have presented two classes of brane solutions. The first one 
represent a class of rotating Dp-branes
in homogeneous plane wave spacetime with both NS-NS and R-R flux. 
We discussed the 
supersymmetry of these branes and their bound states by solving the
type-II Killing spinor equations explicitly. We briefly reviewed the 
possibility of having a regular horizon for this class of branes. 
The worldsheet analysis the background and the rotating branes presented
in section-2 is rather straightforward. 
Following \cite{RT94,RT1}, one
can write down the bosonic sigma model action in the presence of the
non-constant R-R and NS-NS flux. By using the D-brane boundary conditions   
it is easy to write down the mode expansions and thereby the classical
Hamiltonian of the system. So we skip the details here.
Regarding the event horizon, it has been shown in \cite{Brecher1}, in the 
context of Godel background, that one indeed find out a regular horizon
by applying suitable T-duality/dimensional reduction as that removes
the pp-singularities. One could possibly try to find out the
Godel type solutions from the branes solutions presented here and analyze the
properties of horizon, the holographic screens and the 
closed timelike curves. The gauge theory duals of these branes
could also be found out by identifying the supergravity modes with the
corresponding operators in the boundary theory. In the line of the
supergravity theories, an interesting exercise would be to find out
the most general structure of the metric and the other field strengths
and try to solve the field equations. The later would impose
certain constraints of the structure of the flux compatible
with the supersymmetry of the background. 

The second
class of branes were constructed from the nonextremal branes in the
flat spacetime by applying a NMT along the translational isometry
of the brane solutions. This class of branes 
admit regular horizon and the corresponding thermodynamical quantities
like the entropy and temperature are computed. They are found to be 
independent of the parameter $\beta$ that defines the plane wave spacetime.
In summary, the generic solution obtained by null Melvin twist inherit 
their thermodynamic properties from the parent solution in flat spacetime. 
It would be interesting to apply the procedure to construct general 
rotating and charged black brane solutions using this procedure.

\vskip .2in 
\noindent
{\large\bf{Acknowledgment:}} 

\noindent
We would like to thank N. Ohta for useful discussions at the early
stage of the work. We also thank O. Zapata for some comments on the
draft. It's a pleasure to thank
M. Bianchi and A. Sagnotti for encouragement
and for useful discussions. R.R.N. would like to thank String theory
group at the Universita di Roma, `Tor Vergata'
for their kind hospitality where part of the work was done.   
The research of K.P. was supported in part by I.N.F.N., by the E.C. 
RTN programs HPRN-CT-2000-00122 and HPRN-CT-2000-00148, by the 
INTAS contract 99-1-590, by the MURST-COFIN contract 2001-025492 and 
by the NATO contract PST.CLG.978785.

\end{document}